\documentclass[pra,twocolumn,superscriptaddress,showpacs,preprintnumbers,amsmath,amssymb,floatfix]{revtex4-2}
\usepackage{mathtools}
\usepackage{graphicx}% Include figure files
\usepackage{dcolumn}% Alig table columns on decimal point
\usepackage{bm}% bold math
\usepackage{enumerate}
\usepackage{mathdots}
\usepackage[table]{xcolor}
\begin{document}
\title{Phase-sensitive Rydberg-atom interferometry with Floquet electromagnetically induced transparency} 

\author{Yingying Han}
\email[Corresponding email: ]{hanyingying@sztu.edu.cn}
\affiliation{Shenzhen Key Laboratory of  Ultraintense Laser and Advanced Material Technology, Center for Intense Laser Application Technology, and College of Engineering Physics, Shenzhen Technology University, Shenzhen 518118, China}

\author{Changfa He}
\affiliation{Shenzhen Key Laboratory of  Ultraintense Laser and Advanced Material Technology, Center for Intense Laser Application Technology, and College of Engineering Physics, Shenzhen Technology University, Shenzhen 518118, China}
\author{Peng Xu}
\affiliation{School of Physics, Zhengzhou University, Zhengzhou 450001, China}
\affiliation{Institute of Quantum Materials and Physics, Henan Academy of Sciences, Zhengzhou 450046, China}

\author{Yanting Zhao}
\affiliation{State Key Laboratory of Quantum Optics and Quantum Optics Devices, Institute of Laser Spectroscopy,
Shanxi University, Taiyuan 030006, People’s Republic of China}
\affiliation{Collaborative Innovation Center of Extreme Optics, Shanxi University, Taiyuan 030006, People’s Republic of China}

\author{Tao Wang}
\email[Corresponding email: ]{tauwaang@cqu.edu.cn}
\affiliation{Department of Physics, and Center of Quantum Materials and Devices, Chongqing University, Chongqing 401331, China}
\affiliation{Center of Modern Physics, Institute for Smart City of Chongqing University in Liyang, Liyang 213300, China}

\author{Weidong Li}
\email[Corresponding email: ]{liweidong@sztu.edu.cn}
\affiliation{Shenzhen Key Laboratory of  Ultraintense Laser and Advanced Material Technology, Center for Intense Laser Application Technology, and College of Engineering Physics, Shenzhen Technology University, Shenzhen 518118, China}

\date{\today}

\begin{abstract}
We design a phase-sensitive Rydberg-atom interferometry by implementing Floquet electromagnetically induced transparency (FEIT). The FEIT mixes the sidebands of a Rydberg state induced by a MHz radio frequency (RF) field and recombines them into FEIT bands. The FEIT bands act as screens to present the interference between paths with different phases, which are transitions between the sidebands and excited states. This interferometry can measure the phase of a MHz RF field without a local RF reference field. A phase reference is supplied to the atoms via a periodic electrical signal in the FEIT. We demonstrate that the MHz RF phase is measured over a full range of $2\pi$, and $10^{-4}$ rad accuracy is achieved. Moreover, the RF amplitude can also be measured with higher accuracy than the traditional EIT-based scheme.
\end{abstract}

\maketitle

\textit{Introduction---}Rydberg-assisted atomic electrometry is an emerging technology that uses transitions between Rydberg states at high principal quantum numbers to detect electric fields ~\cite{WOS001065914200001,WOS000803107000014, WOS:001245518300001}. This is primarily due to the large polarizability offered by high Rydberg states and their ability to act as a calibrated standard for various measurements. In these applications, the electric fields are detected using electromagnetically induced transparency (EIT) both on resonance as Autler–Townes splitting and off-resonance as Stark shifts~\cite{PhysRevA.100.063427,WOS:001096527500024}. For the measurements of radio frequency (RF) fields, Rydberg atoms have been used to extend the frequency range from MHz to THz~\cite{PhysRevLett.98.113003,WOS:000822616200003,10.1063/5.0162101,WOS:000345849100025} and improve the sensitivity of RF amplitude~\cite{WOS000310836700021,WOS:000375846600053,PhysRevApplied.18.014045,Yang:24,Yao:22,PhysRevLett.82.1831,2024Continuous} and polarization measurements~\cite{PhysRevLett.111.063001,10.1063/1.5038550,PhysRevApplied.8.014028,PhysRevLett.111.063001}, but less schemes for RF phase measurements.

Interferometric phase sensing is often performed by linear superposition of RF signal and local RF reference
waves with homodyne or heterodyne principle~\cite{WOS:000462086600040,WOS000537039500003}, or closed-loop scheme with optical fields as the reference fields~\cite{2020Rydberg,PhysRevApplied.17.044020,PhysRevApplied.20.054009}. The above methods are suitable for GHz RF fields, which couple the neighboring Rydberg transitions. However, MHz RF fields to couple the neighboring Rydberg transitions require either very high principal quantum number states ($n>100$) or high orbital angular momentum states~\cite{PhysRevResearch.6.023138}, which increase the difficulty of the experiments. Moreover, it is difficult to provide a clean local reference field~\cite{2020Rydberg} and the loop scheme requires a minimum of four fields, resulting in experimental challenges of phase-locking several fields~\cite{WOS:001313607700001}. The MHz RF fields are important for regional broadcasting and aviation air-to-ground communications~\cite{WOS000803107000014,PhysRevApplied.18.014045}, and the Chu's limit of conventional antennas makes it restricted in MHz RF fields measurement~\cite{PhysRevLett.121.110502}. Therefore, it is highly demanded to develop an atomic technique that measures MHz RF phase without local reference field.

In this Letter, we implement the Floquet electromagnetically induced transparency (FEIT) in Rydberg atoms and design a phase-sensitive Rydberg-atom interferometry, in which the quantum interference plays a key role rather than the electromagnetic superposition principle in the Rydberg atom “mixer”~\cite{WOS:000462086600040,WOS000537039500003}. As we know, interferometry is sensitive in relative-phase measurement and is fundamentally important for quantum mechanics, atomic and molecular physics, and precision metrology~\cite{RevModPhys.87.637,PhysRevLett.113.023003,PhysRevLett.122.053601,PhysRevX.2.031011}. This atomic interferometry provides a stand-alone scheme to measure the phase of a MHz RF field without a local RF reference field, and the phase reference is supplied to the atoms via a periodic electrical signal that modulates the control field.

\begin{figure}[thp]
\setlength{\abovecaptionskip}{0.cm}
\setlength{\belowcaptionskip}{-0.cm}
\includegraphics[width=3.4in]{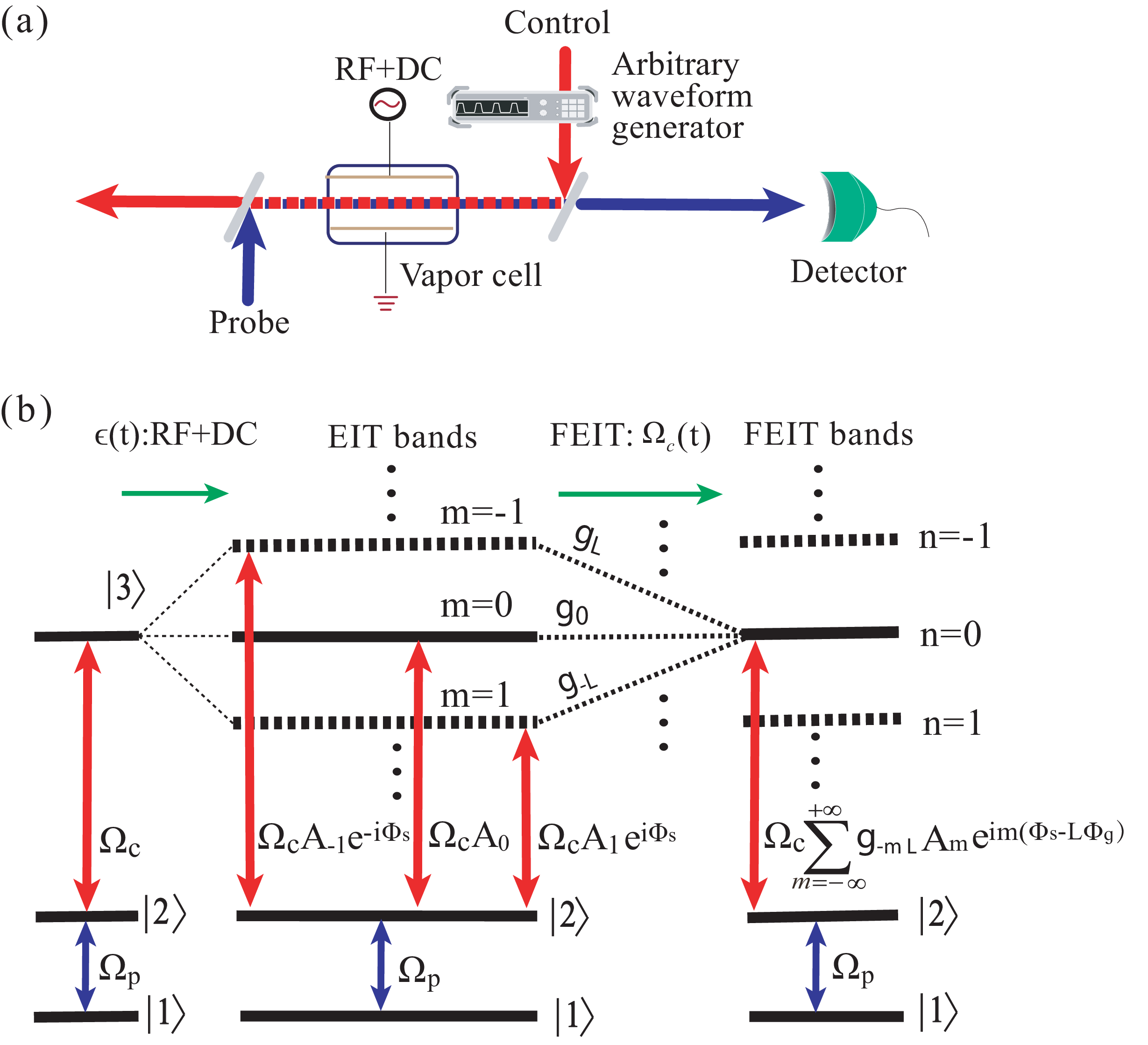}
\caption{\label{fig:1}(a) Schematic of the vapor cell setup with the probe and control beams counter-propagating through the vapor
cell, and the control field modulated by an arbitrary waveform generator. The additional (RF+DC) fields are applied to the two electrode plates. The transmission of the probe beam through the cell is monitored by the detector. (b) Energy-level scheme of the Rydberg-atom interferometry. A three-level system is driven by a resonant probe laser with strength $\Omega_p$ and a detuned control laser with strength $\Omega_c$. The Rydberg state ($|3\rangle$) splits into a series of EIT bands with interval $\omega_s$ under the modulation of a DC and a MHz RF (frequency $\omega_s$, phase $\Phi_s$) field. The EIT bands are tailored with coefficients $g_{n-mL}$ and reassembled into a series of FEIT bands with interval $\omega_g$ under the modulation of FEIT (modulation frequency $\omega_g$). Here, $L=\omega_s/\omega_{g}$ is an integer, and $m$ is the index of the EIT band. RF+DC modulation is similar to a beam splitter and splits the $|2\rangle\leftrightarrow|3\rangle$ transition into infinite paths with different coupling strengths and phases. FEIT with $\Omega_c(t)=\Omega_cg(t)$ is similar to another beam splitter that mixes these paths. $\Phi_g$ is the initial phase of the periodic function $g(t)$.}
\end{figure}

\textit{Multi-channel Rydberg-atom interferometry---}The interferometry is realized in a ladder three-level system consisting of ground state $|1\rangle$ with frequency $\omega_1$, excited state $|2\rangle$ with frequency $\omega_2$ and Rydberg state $|3\rangle$ with frequency $\omega_3$, see Fig.~\ref{fig:1}. A weak probe laser is resonantly coupled to the transition $|1\rangle\leftrightarrow|2\rangle$ with coupling strength $\Omega_p$, and the control laser couples the transition $|2\rangle\leftrightarrow|3\rangle$ with coupling strength $\Omega_c(t)$ for FEIT scheme and $\Omega_c$ for EIT scheme. A weak additional field $\epsilon(t)=\epsilon_{dc}+\epsilon_{rf}\text{cos}(\omega_st+\Phi_s)$ with DC intensity $\epsilon_{dc}$, RF amplitude $\epsilon_{rf}$, to-be-measured phase $\Phi_s$ and frequency $\omega_s$ is applied to the system, and frequency $\omega_s$ with the order of tens of MHz which is far away from the neighboring Rydberg transitions. The Rydberg state is extremely sensitive to the fields, and the energy of the Rydberg state will be $\omega_3(t)=\omega_3-\alpha\epsilon^2(t)/(2\hbar)$ due to the ac Stark effect with $\alpha$ the dipole polarizability of the Rydberg state. 

Firstly, we take a brief look at the EIT scheme with constant $\Omega_c$. In the interaction picture, under conditions $\Omega_c\ll\{\omega_s,\omega_{32}\}$ and $\Omega_p\ll\omega_{21}$ with transition frequencies $\omega_{32}=\omega_{3}-\omega_{2}$ and $\omega_{21}=\omega_{2}-\omega_{1}$, the EIT bands induced by the DC and RF fields with interval $\omega_{s}$ in Fig.~\ref{fig:1}(b) are resolvable~\cite{liu:cpb}. When the control field with frequency $\omega_c$ is close to the $m$-th EIT band, we can safely neglect the other non-resonant terms that oscillate rapidly. Then the $m$-th EIT band (marked as $|m\rangle$), excited and ground states constitute a generalized three-level system, and the Hamiltonian with rotating wave approximation (RWA) is 
$\hat{H}_m^{\text{EIT}}=-\hbar[\Delta_p|2\rangle\langle2|+(\Delta_p+\Delta_m)|m\rangle\langle m|]
-\hbar(\Omega_p|1\rangle\langle2|+\Omega_m^{\text{EIT}}|2\rangle\langle m|+\text{H.c.})/2$, where
\begin{eqnarray}\label{eq:EITomg}
\abovedisplayskip=0pt
\belowdisplayskip=0pt
\Omega_m^{\text{EIT}}=\Omega_cA_{m}e^{im\Phi_s}, 
\end{eqnarray}
with $A_{m}=\sum_{k=-\infty}^{+\infty}J_k\left(\frac{\alpha\varepsilon_{rf}^{2}}{8\hbar\omega_{s}}\right)J_{m-2k}\left(\frac{\alpha\varepsilon_{dc}
\varepsilon_{rf}}{\hbar\omega_{s}}\right)$. Here, $J_k(.)$ is the first kind Bessel function, $\Delta_p=\omega_p-\omega_{21}$, $\Delta_m=\Delta_c+\omega_{\alpha}+m\omega_{s}$ with $\Delta_c=\omega_c-\omega_{32}$, and $\omega_{\alpha}=\alpha\epsilon_{dc}^{2}/(2\hbar)+\alpha\epsilon^{2}_{rf}/(4\hbar)$ is the Stark shift due to the DC and RF fields. From Fig.~\ref{fig:1}(b), we see that the RF+DC modulation splits the Rydberg state into a series of EIT bands~\cite{2009Enhanced,WOS:000377190900004,liu:cpb}, and each EIT band couples the excited state with different coupling strength and RF phase, see Eq.~(\ref{eq:EITomg}). 
Unfortunately, the EIT spectra is proportional to $|\Omega_m^{\text{EIT}}|^2$, which is independent of the RF phase. Therefore, the EIT-based scheme cannot measure the MHz RF phase without a local reference field in this three-level system.

To measure the RF phase in the three-level system with one Rydberg state and without a local RF reference, we modulate the control field with a periodic function $g(t)$ with frequency $\omega_g$ and initial phase $\Phi_g$. The periodic electrical signal can be generated by an arbitrary wave generator or electro-optic modulator, and they are easier to manipulate than the local RF field~\cite{HanPRApplied,2020Rydberg}. This overcomes the difficulty of providing a clean local RF reference field in the previous studies~\cite{2020Rydberg,WOS:000462086600040,WOS000537039500003}. Then the coupling strength of the control field becomes $\Omega_c g(t)$, which is different from the constant $\Omega_c$ in the EIT, and we name this scenario as FEIT. In this FEIT scheme, there are two modulated frequencies: RF frequency $\omega_s$ and frequency $\omega_g$ of the periodic function $g(t)$. Considering the situation that $L=\omega_s/\omega_{g}$ is a positive integer, the EIT bands are reorganized into FEIT bands. Similar to the EIT scheme, with conditions $\Omega_c\ll\{\omega_s,\omega_g,\omega_{32}\}$ and $\Omega_p\ll\omega_{21}$, the FEIT bands in Fig.~\ref{fig:1}(b) with interval $\omega_{g}$ are resolvable, and the interaction Hamiltonian with RWA for the $n$-th FEIT band (marked as $|n\rangle$) is
$\hat{H}_n^{\text{FEIT}}=-\hbar[\Delta_p|2\rangle\langle2|+(\Delta_p+\Delta_n)|n\rangle\langle n|]
-\hbar(\Omega_p|1\rangle\langle2|+\Omega_n^{\text{FEIT}}|2\rangle\langle n|+\text{H.c.})/2$, where
\begin{eqnarray}\label{eq:FEITomg}
\abovedisplayskip=0pt
\belowdisplayskip=0pt
\Omega_n^{\text{FEIT}}=\Omega_{c}\sum_{m}g_{n-mL}A_{m}e^{im(\Phi_s-L\Phi_g)}, 
\end{eqnarray}
details see~\cite{sm}. Here $g_n$ is the Fourier coefficient of the periodic function $g(t)$ and $\Delta_n=\Delta_c+\omega_{\alpha}+n\omega_{g}$. 

From Eqs.~(\ref{eq:FEITomg}) and (\ref{eq:EITomg}), we see that $\Omega_n^{\text{FEIT}}$ is the sum of $\Omega_m^{\text{EIT}}$ with coefficients $g_{n-mL}$ and relative phase $\Phi_s-L\Phi_g$. Note that $L$ is constant for a determined $\omega_g$ and $\omega_s$. Thus the FEIT modulation is similar to the second beam splitter in the Mach-Zehnder interferometer~\cite{MZIscirep,PhysRevApplied.19.064021,Li:24}, which mixes the paths between the EIT bands and the excited state. The FEIT spectra is proportional to $|\Omega_n^{\text{FEIT}}|^2$, where the cross terms represent the interference between different paths, and the accumulated phase along the $m$-th path is $m(\Phi_s-L\Phi_g)$. Then the FEIT bands serve as screens to present the quantum interference. Moreover, we can design the interference paths and the accumulated phases through Floquet engineering, i.e., changing the periodic function $g(t)$. For example, for the cosine function $g(t)=[1+\text{cos}(\omega_g t+\Phi_g)]/2$ with $\omega_g=\omega_s$ (i.e., $L=1$), we set the 1st FEIT band as the screen, then $|\Omega_1^{\text{FEIT}}|^2=\left|\Omega_{c}(A_{0}/4+A_{1}e^{i\Delta_{\Phi}}/2+A_{2}e^{2i\Delta_{\Phi}}/4)\right|^2$ with $\Delta_{\Phi}=\Phi_s-\Phi_g$, which indicates the interference among paths $|2\rangle\leftrightarrow|m=0\rangle$ with accumulated phase 0, $|2\rangle\leftrightarrow|m=1\rangle$ with accumulated phase $\Delta_{\Phi}$ and $|2\rangle\leftrightarrow|m=2\rangle$ with accumulated phase $2\Delta_{\Phi}$. The FEIT proposed here are readily realized in many three-level systems, such as atomic gases, artificial atoms in superconducting quantum circuits~\cite{Han:22,PhysRevA.101.022108,You2011Atomic}, and three-level meta-atoms in meta-materials~\cite{PhysRevLett.101.253903,PhysRevLett.107.043901}.

To study the Rydberg spectra commonly observed in experiments, we investigate the susceptibility $\chi$, the imaginary part of which indicates the absorption characteristic of the medium, and consider the Doppler effect due to the non-zero velocity distribution of atoms in the medium by integrating over velocity $\nu$ 
\begin{small}
\begin{align}\label{eq:kai1}
\abovedisplayskip=0pt
\belowdisplayskip=0pt
	\chi=\int \frac{iN(\nu)|\mu_{12}|^2/\hbar\varepsilon_0}{\Gamma_{1}-i\Delta_{p}-i\omega_p\frac{\nu}{c}+\frac{|\Omega_{n}^{\text{FEIT}}|^{2}/4}{\Gamma_{2}-i(\Delta_{p}+
\Delta_{n})-i(\omega_{p}-\omega_{c})\frac{\nu}{c}}}d\nu,
\end{align}
\end{small}
where $N(\nu)=N_0\text{exp}(-\nu^2/u^2)/(u\sqrt{\pi})$ with $N_0$ the atom number density in the cell, $u=\sqrt{2k_{b}T_{0}/m}$ the most probable velocity with atom mass $m$, Boltzmann constant $k_{b}$ and medium temperature $\text{T}_{0}$, $c$ the speed of light, $\varepsilon_0$ the permittivity of vacuum, $\mu_{12}$ the dipole matrix element corresponding to the transition $|1\rangle\leftrightarrow|2\rangle$, $\Gamma_{1,2}$ the decay parameters of the system. Then the total transmission is 
%\begin{small}
\begin{align}\label{eq:kai}
\abovedisplayskip=0pt
\belowdisplayskip=0pt
	\text{T}^{\text{Tot}}=\sum_{n}\text{T}_{n}=\sum_{n}\mathrm{exp}\left[-2\pi l\text{Im}(\chi)/\lambda_{p}\right]
\end{align}
%\end{small}
with the $n$-th sideband transmission $\text{T}_{n}$, vapor cell length $l$ and probe field wavelength $\lambda_{p}$.

\begin{figure}[thp]
\setlength{\abovecaptionskip}{0.cm}
\setlength{\belowcaptionskip}{-0.cm}
\includegraphics[width=3.6in]{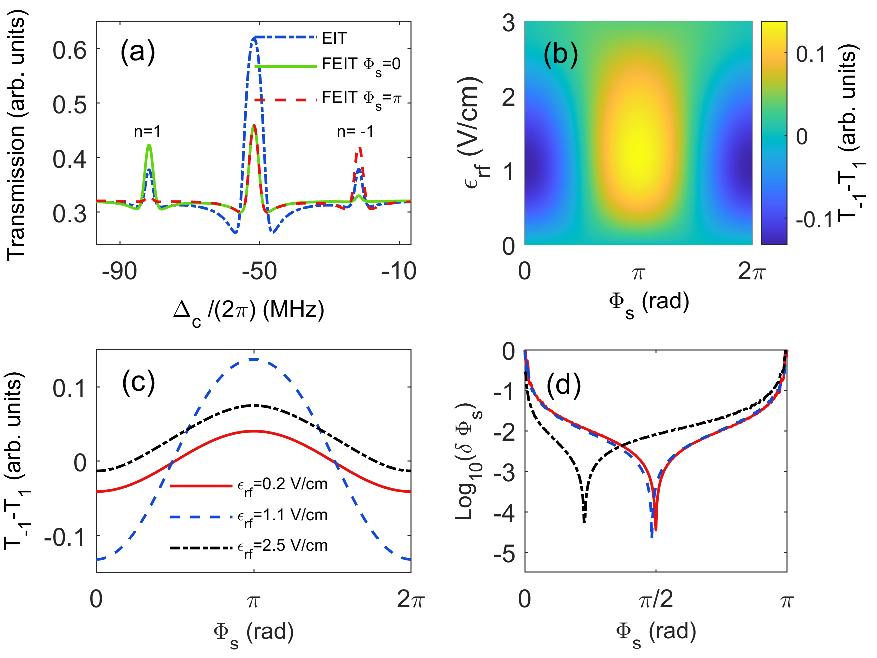}
\caption{\label{fig:2}(a) The EIT and FEIT spectra with different RF phase under $\epsilon_{dc}=3$ V/cm, $\epsilon_{rf}=0.5$ V/cm and in this figure $\Phi_g=0$. (b) The color plot of $\text{T}_{-1}-\text{T}_{1}$ as a function of RF phase $\Phi_s$ and amplitude $\epsilon_{rf}$ with $\epsilon_{dc}=3$ V/cm. Here the results are calculated by repeating Eq.~(\ref{eq:kai}) with $\Omega_n^{\text{FEIT}}$ in Eq.~(\ref{eq:FEITomg}) 10,000 times while taking into account $1\%$ fluctuation of $\Omega_c$ and 100kHz fluctuations of probe frequency $\omega_p$ and control frequency $\omega_c$. Because $\Omega_p$ is very small, we ignore its fluctuation. The results shown in (b) are the average values. (c) $\text{T}_{-1}-\text{T}_{1}$ as a function of RF phase ($\Phi_s$) with $\epsilon_{rf}=0.2$ V/cm (red solid line), $\epsilon_{rf}=1.1$ V/cm (blue dashed line) and $\epsilon_{rf}=2.5$ V/cm (black dash-dotted line). (d) The accuracy $\delta{\Phi_s}$ as a function of RF phase. The accuracy defined as $\delta \Phi_s=\delta (\text{T}_{-1}-\text{T}_{1})/[\partial (\text{T}_{-1}-\text{T}_{1})/\partial\Phi_s]$ with standard deviation $\delta (\text{T}_{-1}-\text{T}_{1})$.}
\end{figure}
\begin{figure}[thp]
\setlength{\abovecaptionskip}{0.cm}
\setlength{\belowcaptionskip}{-0.cm}
\includegraphics[width=2.6in]{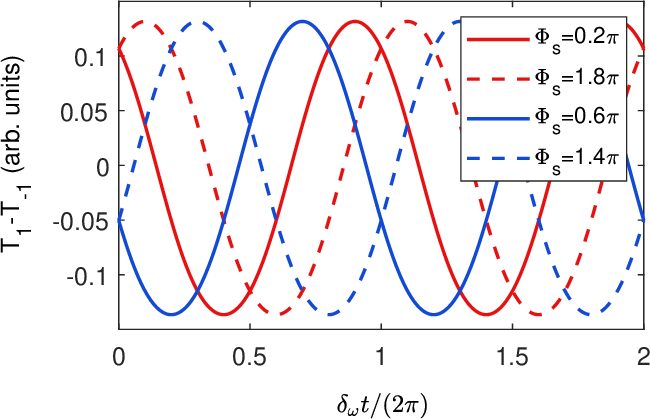}
\caption{\label{fig:3}The oscillations of $\text{T}_{1}-\text{T}_{-1}$ with time under different RF phase with $\epsilon_{rf}=1.1$ V/cm, $\delta_\omega/(2\pi)=0.005~\text{MHz}$, $\epsilon_{dc}=3$ V/cm and $\Phi_g=0$. The results are calculated by Eq.~(\ref{eq:kai}) with $\Omega_n^{\text{FEIT}}(t)$ in Eq.~(\ref{eq:FEITomgt}) and $L=1$.}
\end{figure}
\textit{Megahertz RF phase measurement---}We present a specific scheme to measure the phase of a MHz RF field. The parameters of the three-level system are shown in Ref.~\cite{not}. A commonly used cosine function is considered: $g(t)=[1+\text{cos}(\omega_gt+\Phi_g)]/2$, and without loss of generality, we set the reference phase to 0, i.e., $\Phi_g=0$. 

In Fig.~\ref{fig:2}(a), we show the EIT spectra with coupling strength in Eq.~(\ref{eq:EITomg}), and the RF frequency can be obtained by measuring the interval between the adjacent bands. Then we set the modulation frequency of FEIT equal to the RF frequency, i.e., $\omega_g/(2\pi)=\omega_{s}/(2\pi)=30~\text{MHz}$. In Fig.~\ref{fig:2}(a), we also show the FEIT spectra with coupling strength in Eq.~(\ref{eq:FEITomg}) under different RF phases. We see that the FEIT spectra vary with RF phase and the main peak is shifted by the Stark shift $\omega_{\alpha}/(2\pi)\approx-51.6~\text{MHz}$. Since $|\Omega_m^{\text{EIT}}|^2$ is phase independent, the EIT spectra do not change with the RF phase. For the FEIT spectra, the maximum transmissions of the first-order bands (marked by $\text{T}_{n=\pm1}^{\text{max}}$, $\text{T}_{\pm1}$ for short) change in opposite with RF phase, i.e., when the RF phase changes from 0 to $\pi$, $\text{T}_{1}$ decreases and $\text{T}_{-1}$ increases. The variation of $\text{T}_{\pm1}$ with RF phase is similar to the variation of stripe brightness with the phase difference of the paths in interference experiments. To increase the contrast of spectral amplitudes and eliminate the same fluctuations on the $n=1$ and $n=-1$ bands, we show $\text{T}_{-1}-\text{T}_{1}$ varies with RF phase and RF amplitude in Fig.~\ref{fig:2}(b). We find that the contrast $\text{T}_{-1}-\text{T}_{1}$ is monotonically increasing with $\Phi_s$ in the range of $[0,\pi]$, and monotonically decreasing with $\Phi_s$ in the range of $[\pi,2\pi]$. More details are shown in Fig.~\ref{fig:2}(c), which show the contrasts vary with RF phase with specific $\epsilon_{rf}$. We find that these curves are symmetric about $\pi$, which leads to the inability to distinguish $\Phi_s\in[0,\pi]$ and $2\pi-\Phi_s$. We will discuss how to distinguish them later. The accuracies $\delta{\Phi_s}$ over a range of $[0,\pi]$ are shown in Fig.~\ref{fig:2}(d). We find that the minimum detectable $\delta{\Phi_s}\approx10^{-4}$ rad. The accuracy curves within the range of $[\pi,2\pi]$ are similar to that in $[0,\pi]$.

To distinguish $\Phi_s\in[0,\pi]$ and $2\pi-\Phi_s$, we consider the situations where the modulation frequency $\omega_g$ is close to the RF frequency $\omega_s$ with a small detuning, i.e., $\delta_\omega=\omega_s-\omega_g\ll\{\omega_s,\omega_g\}$, and $\delta_\omega$ is much smaller than other parameters in the system. Then Eq.~(\ref{eq:FEITomg}) becomes
\begin{eqnarray}\label{eq:FEITomgt}
\abovedisplayskip=0pt
\belowdisplayskip=0pt
 \Omega_n^{\text{FEIT}}(t)=\Omega_{c}\sum_{m}g_{n-mL}A_{m}e^{im(\delta_\omega t+\Phi_s-L\Phi_g)},
 \end{eqnarray}
details see~\cite{sm}. The FEIT spectra will vary over time on the time scale of $2\pi/\delta_{\omega}$. In Fig.~\ref{fig:3}, we show the oscillations of $\text{T}_{1}-\text{T}_{-1}$ with time under different RF phase. We see that although $\text{T}_{1}-\text{T}_{-1}$ have the same values at the initial time for $\Phi_s=0.2\pi$ and $\Phi_s=1.8\pi$, their curves change completely differently over time. Therefore, we can distinguish $\Phi_s\in[0,\pi]$ and $2\pi-\Phi_s$ based on the curves of $\text{T}_{1}-\text{T}_{-1}$ changing over time. The analysis of the cosine-like curves in Fig.~\ref{fig:3} is presented in Ref.~\cite{sm}. In summary, first set $\omega_{g}=\omega_s$, we can determine the phase $\Phi_s\in[0,\pi]$ or $2\pi-\Phi_s$, then set small $\delta_\omega=\omega_s-\omega_g$, according to the curve of $\text{T}_{1}-\text{T}_{-1}$ over time to distinguish $\Phi_s\in[0,\pi]$ and $2\pi-\Phi_s$.

\begin{figure}[thp]
\setlength{\abovecaptionskip}{0.cm}
\setlength{\belowcaptionskip}{-0.cm}
\includegraphics[width=3.4in]{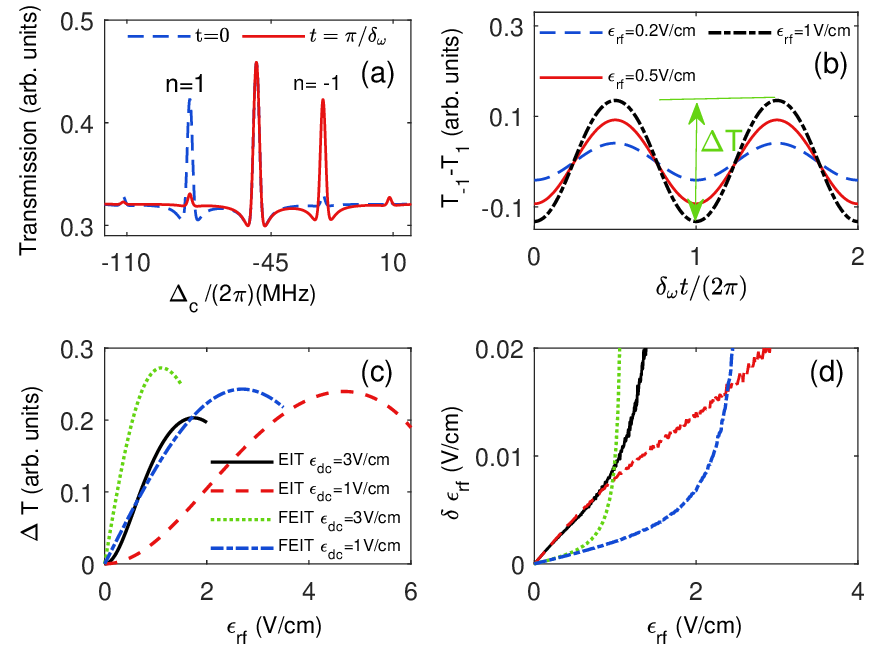}
\caption{\label{fig:4}(a) The FEIT spectra at different times: $t=0$ for the blue dashed line and $t=\pi/\delta_\omega$ for the red solid line with $\epsilon_{dc}=3$ V/cm, $\epsilon_{rf}=0.5~\text{V/cm}$. In this figure $\Phi_s=\Phi_g=0$ and $\delta_\omega/(2\pi)=0.005~\text{MHz}$. (b) The oscillations of $\text{T}_{-1}-\text{T}_{1}$ with time under different RF amplitude with $\epsilon_{dc}=3$ V/cm. (c) The contrasts $\Delta T$ of the oscillations in (b) vary with $\epsilon_{rf}$ for different cases. (d) The accuracies $\delta_{\epsilon_{rf}}$ as a function of $\epsilon_{rf}$ for the cases in (c). Here the FEIT results are calculated by Eq.~(\ref{eq:kai}) with $\Omega_n^{\text{FEIT}}(t)$ in Eq.~(\ref{eq:FEITomgt}), and the accuracies by considering the same noises as in Fig.~\ref{fig:2}. The EIT results are calculated by Eq.~(\ref{eq:kai}) with $\Omega_n^{\text{EIT}}$ in Eq.~(\ref{eq:EITomg}). The accuracy defined as $\delta \epsilon_{rf}=\delta (\Delta T)/[\partial (\Delta T)/\partial\epsilon_{rf}]$ with standard deviation $\delta (\Delta T)$.}
\end{figure}
\textit{Megahertz RF amplitude measurement---}Here is another specific example of measuring the RF amplitude. In this section, we also consider the situations that there is a small detuning between $\omega_s$ and $\omega_{g}$, and the coupling strength between the $n$-th FEIT band and the excited state is Eq.~(\ref{eq:FEITomgt}). Because $\delta_\omega$ is much smaller than other system energy scales, we can treat the $|\Omega_n^{\text{FEIT}}(t)|^2$ as a constant-like term during the process of obtaining the Rydberg spectra. 

In Fig.~\ref{fig:4}(a) we show the FEIT spectra at different times. It is shown that the first-order bands ($n=\pm1$) indeed change most dramatically with time. We lock the control detuning at the peak of the $n=\pm1$ bands and measure the maximum transmissions of them over time scale of $2\pi/\delta_\omega$, respectively. Then the oscillations of $\text{T}_{-1}-\text{T}_{1}$ under different RF amplitude are shown in Fig.~\ref{fig:4}(b). We see that the frequencies of the oscillates are $\delta_{\omega}$, and the contrasts of the oscillation increase with the RF amplitude. Here the contrast $\Delta\text{T}$ for the FEIT scheme is defined as the difference between the maximum and the minimum amplitudes of the oscillation in Fig.~\ref{fig:4}(b), and the contrast $\Delta\text{T}$ for the EIT scheme is $\Delta \text{T}(\epsilon_{rf})=\text{T}_{-1}|_{\epsilon_{rf}}-\text{T}_{-1}|_{\epsilon_{rf}=0}$. The contrasts vary with RF amplitude for the EIT and FEIT schemes are shown in Fig.~\ref{fig:4}(c). We see that the contrasts are
monotonically increasing with $\epsilon_{rf}$ until $\epsilon_{rf}\simeq\{1.1~\text{V/cm}, 2.7~\text{V/cm}\}$ for FEIT scheme and $\epsilon_{rf}\simeq\{1.7~\text{V/cm}, 4.6~\text{V/cm}\}$ for EIT scheme for $\epsilon_{dc}=\{3~\text{V/cm}, 1~\text{V/cm}\}$, respectively. The measurement range of RF amplitude can be adjusted by changing the DC intensity. The corresponding accuracies are shown in Fig.~\ref{fig:4}(d), we see that the minimum distinguishabilities of the FEIT scheme in the ranges $\epsilon_{rf}\in(0,1)$ and $(0,2.4)$ are smaller than that of the EIT scheme for $\epsilon_{dc}=3~\text{V/cm}$ and $1~\text{V/cm}$, respectively. Therefore, compared with the EIT-based scheme, the FEIT-based scheme shows the advantage of high accuracy in measuring the amplitude of weak RF fields under the same parameter conditions.

\textit{Conclusion---}During the last few decades, EIT has been widely used in Rydberg atoms. Here we implement the FEIT with a periodically modulated control field and propose a phase-sensitive Rydberg-atom interferometry to measure the phase and amplitude of a MHz RF field. We demonstrate that the RF phase is measured over a full range of $2\pi$ without a local RF reference field and the RF amplitude is measured with a higher accuracy than the EIT scheme. The Rydberg-atom interferometry induced by FEIT can be goal-oriented designed by Floquet engineering, i.e., changing the frequency, phase and amplitude of the periodic function. For future investigations, this interferometry is promises to advance interference-based electric-field sensors, such as measurement of the GHz RF phase, continuous frequency band, the polarization direction and intensity of DC field, and so on.

\textit{Acknowledgements---}We thank Augusto Smerzi and Wenxian Zhang for fruitful discussions. This work is supported by the National Natural Science Foundation of China (Grant No.~12205199, No. 12375023, No. 12204428, No. 12274272, No. 12274045 and No. 12347101), National Key Research
and Development Program of China (Grants No. 2022YFA1404201), Natural Science Foundation of Top Talent of SZTU (Grant No. GDRC202202 and GDRC202312), the Natural Science Foundation of Henan Province (Grant No.242300421159) and Guangdong Provincial Quantum Science Strategic Initiative (No.GDZX2305006).

%\bibliography{ref}

\begin{thebibliography}{45}
\bibitem[{not()}]{not}
\bibinfo{note}{Considering a $\prescript{133}{}{\text{Cs}}$ three-level system:
  $6S_{1/2}-6P_{3/2}-53S_{1/2}$ in a vapor cell at a controlled temperature of
  $300$K. The polarizability of the Rydberg state $53S_{1/2}$ is
  $\alpha=71.1\hbar~\text{MHz}/(\text{V/cm})^2$. We set the decay parameters
  $\Gamma_1/(2\pi)=2.65~\text{MHz}$ and $\Gamma_2/(2\pi)=0.015~\text{MHz}$, the
  coupling Rabi frequency $\Omega_c/(2\pi)=5~\text{MHz}$, the probe Rabi
  frequency $\Omega_p/(2\pi)=0.1~\text{MHz}$ and the probe dutuning
  $\Delta_p=0$ in this work.}
\end{thebibliography}

\end{document}